\documentclass[a4paper, pra, aps, twocolumn, 10pt]{revtex4-2}

\usepackage[utf8x]{inputenc}
\usepackage{amsmath}
\usepackage{amsfonts}
\usepackage{textcomp}
\usepackage{amssymb}
\usepackage{empheq}
\usepackage{esvect}
\usepackage{subcaption}

\usepackage{xcolor}
\definecolor{lightpink}{RGB}{255, 230, 230}
\usepackage{hyperref}
\hypersetup{
  colorlinks   = true, 
  urlcolor     = blue, 
  linkcolor    = blue, 
  citecolor   = red 
}


\begin{document}
\author{Nikita~A.~Nemkov}
\email{nnemkov@gmail.com}
\affiliation{Sber Quantum Technology Center, Kutuzovski prospect 32, Moscow, 121170, Russia}

\author{Stanislav~S.~Straupe}
\affiliation{Sber Quantum Technology Center, Kutuzovski prospect 32, Moscow, 121170, Russia}

\title{Complexity-energy trade-off in programmable unitary interferometers}
\begin{abstract}
Coherent multiport interferometers are a promising approach to realize matrix multiplication in integrated photonics. However, most known architectures -- such as MZI and beamsplitter meshes, as well as more general interferometers -- suffer from complicated procedures for mapping the matrix elements of the desired transformation to specific phaseshifts in the device. We point out that the high programming complexity is intrinsic, rather than accidental. At the same time, we argue that interferometers admitting efficient programming algorithms in general yield a much lower useful output energy, which ultimately limits their accuracy and energy efficiency.
\end{abstract}
\maketitle

\section{Introduction}

Photonic hardware has emerged as a promising platform for accelerating linear algebra, a core bottleneck in modern machine learning workloads. Optical systems can exploit the inherent parallelism and high bandwidth of light to perform matrix-vector multiplications with low latency and power consumption \cite{McMahon2023,shastriPhotonicsDeepLearning2021, feldmannParallelProgrammingOptical2021,zhou2022photonic}. In particular, integrated optical circuits with programmable interferometers have been proposed as analog accelerators for deep learning inference, promising much higher throughput, lower latency, and extreme energy efficiency compared to their electronic counterparts \cite{OuScience2025,XuScience2024}.

Beyond machine learning, programmable optical circuits also play a central role in quantum information processing \cite{wang2020integrated}, where they enable reconfigurable linear optical networks for boson sampling \cite{hoch2022reconfigurable,hoch2025quantum} and universal quantum computing \cite{wangMultidimensionalQuantumEntanglement2018,MaringNaturePhotonics2024,psiquantum2025manufacturable}.

Reconfigurable multiport unitary interferometers are key components in photonic systems for quantum computing and optical neural networks \cite{wu2024programmable,kim2025programmable}. Many engineering challenges must be overcome to make large-scale integrated interferometers possible. In particular, characterization, calibration, and stabilization are all rather non-trivial \cite{HamerlyPRApplied2022,HamerlyPRApplied2022,bandyopadhyay2021hardware,FldzhyanOptLett2020}. Here, we will ignore these practical issues and focus on the programming algorithms for idealized unitary interferometers.

Importantly, programming (configuring) multiport interferometers often has high computational cost. For instance, it is customary to implement a generic linear transformation through the singular value decomposition (SVD) on a chip \cite{clementsOptimalDesignUniversal2016a}. Computing the SVD decomposition for an $n\times n$ matrix, however, has complexity $O(n^3)$ \cite{golub13}. In the context of accelerating linear algebra, this substantial fixed cost may negate the advantages of using analog optical computing.

In this work, we argue that this high programming complexity is quite universal and shared by many reconfigurable interferometers, including those not relying on SVD. At the same time, we note that easy-to-program interferometers are in general suboptimal with respect to how much useful output energy they produce, which can limit their accuracy and energy efficiency.

\section{Programming as embedding}
\begin{figure}
    \centering
    \includegraphics[width=\linewidth]{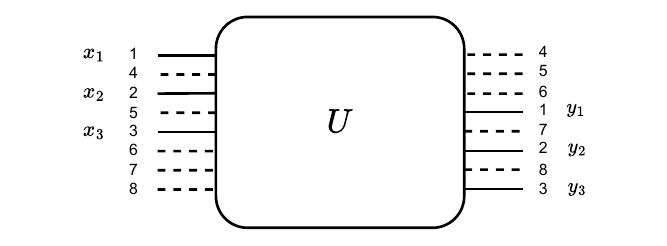}
    \caption{A sketch of a unitary interferometer with 3+5 input and 3+5 output channels. Dashed lines denote auxiliary ports.}
    \label{fig U}
\end{figure}

Consider a generic unitary interferometer with $n+m$ input and output channels, see Fig.~\ref{fig U} for a sketch. We will say that the interferometer implements an $n\times n$ matrix $A$ if it is configured so that $n$ of its input amplitudes $x_1,\dots, x_n$ are related to $n$ of its output amplitudes $y_1,\dots y_n$ by a linear transformation $y = Ax$. The remaining $m$ input and output ports are auxiliary. If we label the ports of the interferometer so that $x_{1-n}$ and $y_{1-n}$ map to the first $n$ input and output ports, respectively, the unitary matrix of the interferometer has the following block structure
\begin{align}
    U = \begin{pmatrix}A & B \\ C & D \end{pmatrix} \ , \label{embedding}
\end{align}
i.e. $A$ is embedded in $U$ as the top left block. This is the simplest example of such an embedding, and other realizations are known \cite{TangLowDepth,Fldzyan2024LowDepth,Fldzhyan2025lowdepth,Hamerly2024}, which are advantageous in some practical cases; however, our arguments in the following are general and do not depend on a particular embedding.

Not every matrix can be embedded in a unitary. In App.~\ref{app unitarity} we show that as long as the number of auxiliary modes $m$ is sufficiently large $m\ge n$, an arbitrary matrix $A$ can be embedded if and only if its largest singular value $\sigma_{\max} (A)$ is less or equal to one 
\begin{align}
 \sigma_{\max}(A)\le 1   \ .
\end{align}
This condition has a clear physical interpretation. Indeed, the maximum singular value, also known as the spectral norm $\|A\|_2$, is equal to 
\begin{align}
    \sigma_{\max}(A) = \max_{x} \frac{\|Ax\|^2}{\|x\|^2} \ ,
\end{align}
where $\|\cdot\|$ is the standard Euclidean vector norm. If $x$ is a vector of amplitudes, $\| x\|^2$ is the total energy in the corresponding ports. When amplitudes of the $m$ auxiliary input ports are zero, $\sigma_{\max}(A)$ is the maximum ratio of the output energy (in the first $n$ ports) to the total input energy, which can not exceed one in a unitary interferometer.

\section{Programming generic interferometers}
\begin{figure}
    \centering
    \begin{subfigure}{\linewidth}
    \includegraphics[width=\linewidth]{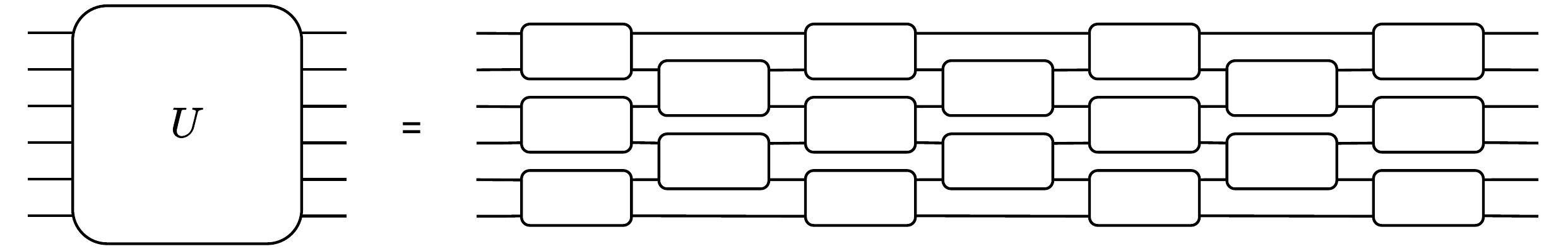}
    \caption{A mesh of MZI implementing $6\times 6$ Clements interferometer.}
    \label{fig Clements}
    \end{subfigure}

    \begin{subfigure}{\linewidth}
    \includegraphics[width=\linewidth]{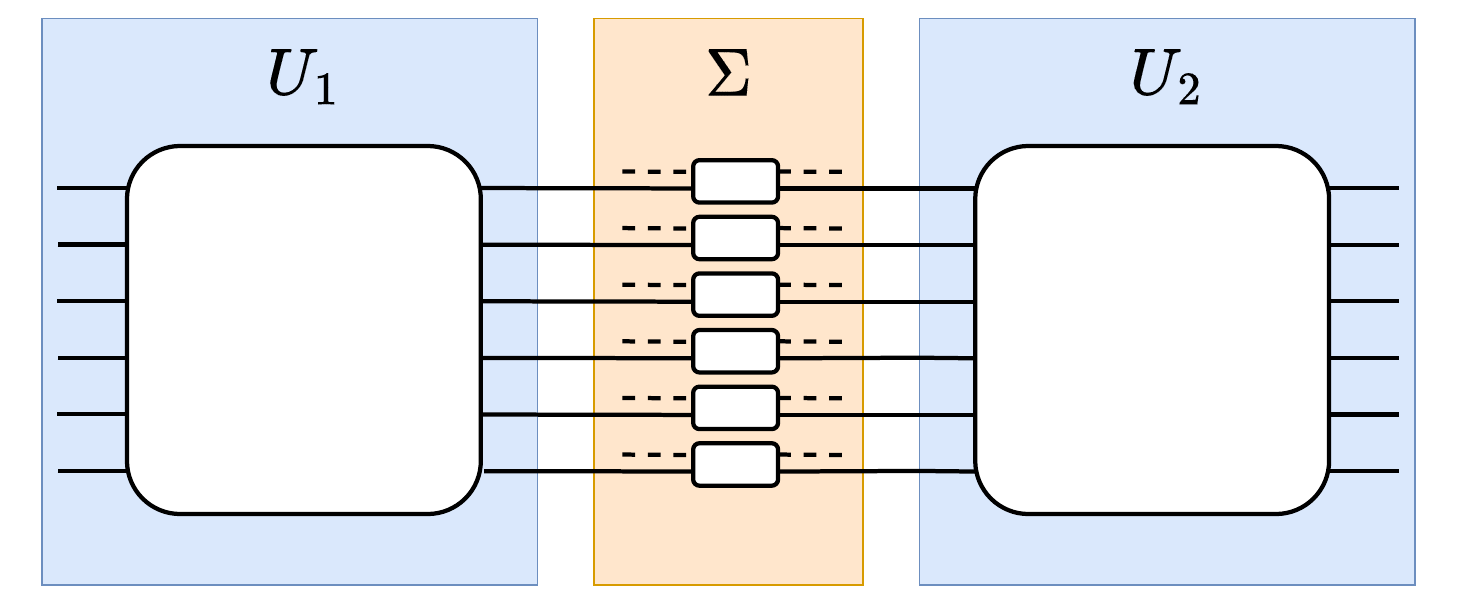}
    \caption{Arbitrary matrix may be embedded into a larger unitary interferometer using SVD decomposition.}
    \label{fig svd}
    \end{subfigure}
    \caption{Realizing unitary and arbitrary transformations with coherent interferometers.}
\end{figure}

Some unitary interferometers can implement arbitrary matrices $A$ subject to $\sigma_{\max}(A)\le 1$. Perhaps, the most familiar examples are architectures based on the Reck \cite{reckExperimentalRealizationAny1994} and Clements \cite{clementsOptimalDesignUniversal2016a} meshes of Mach-Zehnder interferometers (MZIs). For example, the Clements mesh (see Fig.~\ref{fig Clements}) can be configured to implement an arbitrary unitary matrix $U$. To obtain a general non-unitary matrix $A$ one uses the singular value decomposition (SVD) 
\begin{align}
A=U_1 \Sigma U_2 \ , \label{svd}    
\end{align}
which represents $A$ as a product of two unitaries $U_1, U_2$ sandwiching a diagonal matrix of singular values $\Sigma$. This diagonal matrix is the only non-unitary part, and can be realized e.g. using a column of MZIs with one auxiliary input and one output port per optical channel, see Fig.~\ref{fig svd}.

One drawback of this scheme is the substantial numerical cost of performing the SVD. Indeed, computing the exact SVD of a dense $n\times n$ matrix has $O(n^3)$ complexity \cite{golub13}. The main observation of this work is that this high cost is not unique to the Clements-based implementation (which explicitly uses SVD) but much more general.

\begin{figure}
    \centering
    \includegraphics[width=\linewidth]{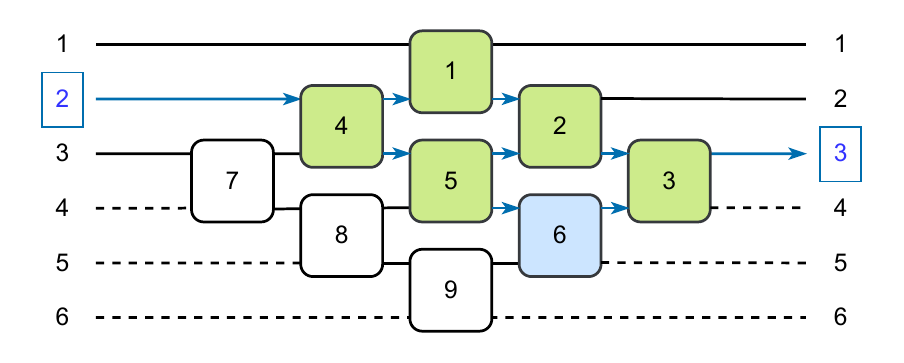}
    \caption{Diamond mesh. Blue lines denote optical paths that contribute to the matrix element $A_{23}$.}
    \label{fig diamond}
\end{figure}

For a concrete example, consider a diamond mesh \cite{Hamerly2024} shown in Fig.~\ref{fig diamond}. This mesh has $m=n$ auxiliary ports and there is a straightforward algorithm to program it, though its complexity is not obvious. For the $3\times 3$ example depicted, there are 9 MZI blocks which can be programmed consecutively in the order specified in the figure. For instance, the parameters of the first MZI block can be set based on the value of matrix element $A_{11}$ alone, because there is a single optical path connecting the first input and the first output ports, and it only goes through the first MZI. When the first MZI is set, the parameters of the second MZI can be determined from the matrix element $A_{12}$, and so forth. Now assume that MZI blocks 1 to 5 have been configured, and consider setting block 6. This block can be configured based on matrix element $A_{23}$. Though this is straightforward in principle, multiple optical paths now contribute, so the number of numerical operations needed to determine the parameters of MZI block 6 is non-trivial. Apparently, the numerical complexity of configuring this interferometer increases faster than $O(n^2)$, because the number of operations per MZI block also scales with $n$.

We believe that these two rather different examples illustrate a general principle. Consider a programming algorithm that can embed an arbitrary matrix $A$ in a unitary interferometer as long as $\sigma_{\max}(A)\le 1$. Such algorithm, if applied to a matrix with $\sigma_{\max}>1$, must fail. Therefore, any such algorithm performs what we call a \textit{singular value test} (SVT) as a byproduct, i.e. it can determine whether $\sigma_{\max}(A)\le1$ (completes correctly) or $\sigma_{\max}(A)>1$ (signals an error). Therefore, the complexity of such an algorithm is at least the complexity of the SVT.

To the best of our knowledge, the complexity of SVT has not been rigorously established. While there are probabilistic \cite{Hochstenbach2013, VanDorsselaer2000} or approximate \cite{golub13, Yousef} algorithms with complexity $O(n^2)$, all exact deterministic algorithms are likely bounded by the matrix multiplication complexity $O(2^\omega)$ \cite{STRASSEN1969} in theory and the full SVD cost $O(n^3)$ in practice. We view this as strong evidence that the high complexity of known programming algorithms for unitary interferometers is intrinsic and universal.

\section{Programming restricted interferometers and complexity-energy trade-off}
\begin{figure}
    \centering
    \begin{subfigure}{\linewidth}
    \includegraphics[width=\linewidth]{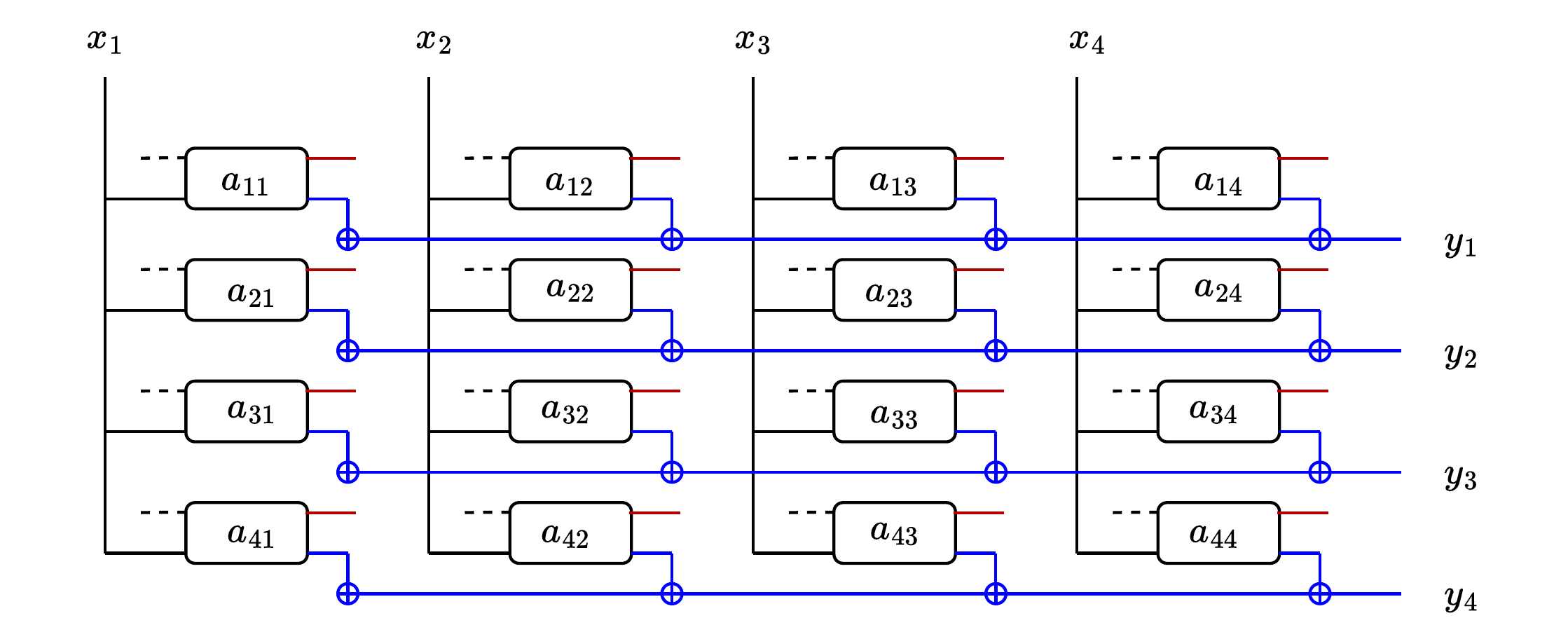}
    \caption{Standard coherent crossbar architecture. Red color highlights ports contributing to energy loss.}
    \label{fig xbar1}
    \end{subfigure}

    \begin{subfigure}{\linewidth}
    \includegraphics[width=\linewidth]{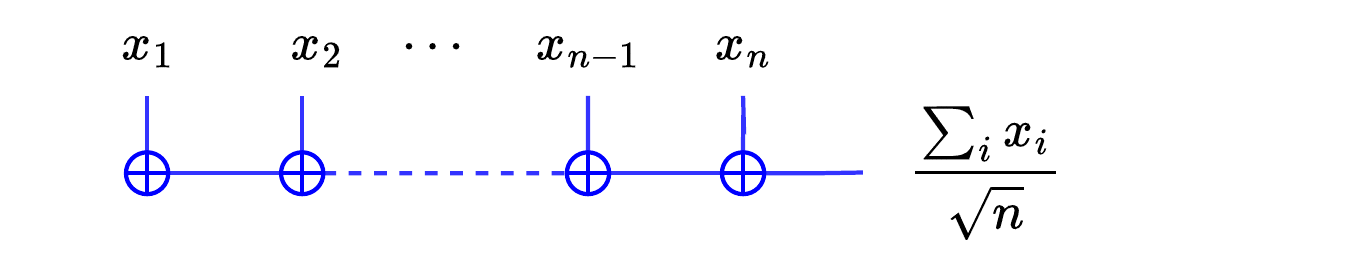}
    \caption{A unitary optical summator (fan-in).}
    \label{fig fanin}
    \end{subfigure}

    \begin{subfigure}{\linewidth}
    \includegraphics[width=\linewidth]{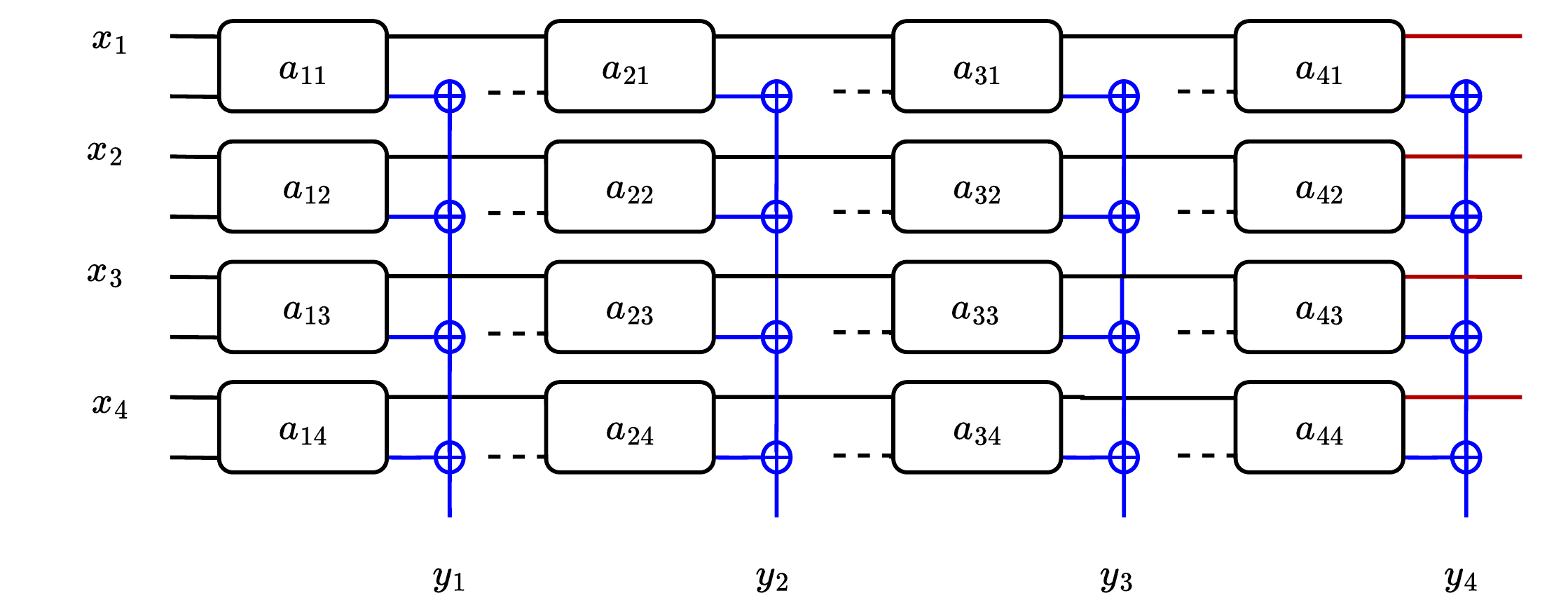}
    \caption{Modified coherent crossbar architecture. Red color highlights ports contributing to energy loss.}
    \label{fig xbar2}
    \end{subfigure}
    
    \caption{Coherent optical crossbar arrays illustrating the general complexity-energy trade-off in unitary interferometers.}
\end{figure}

Some unitary interferometers have low, merely quadratic programming cost. One class of examples are various optical crossbar arrays \cite{Zou2022,Feldmann2021,Xu2021b,Giamougiannis2022a,Sturm2022,Giamougiannis2022}. A version of a coherent unitary crossbar is schematically depicted in Fig.~\ref{fig xbar1}. In this case, there is a direct mapping between the elements of the matrix to be embedded $W$ and the individual optical elements. This does not contradict the SVT argument for a somewhat subtle reason, that we now explain.

The interferometers of the previous section can embed an arbitrary matrix $W$ rescaled by its maximum singular value $A=\frac{W}{\sigma_{\max}(W)}$. Similarly, interferometer Fig.~\ref{fig xbar1} can also implement an arbitrary matrix $W$, but with a different rescaling 
\begin{align}
 A=\frac{W}{\alpha_{\max}(W)},\quad \alpha_{\max}(W)=n \max_{ij}|W_{ij}| \ .
\end{align}

Let us unpack that. Because each MZI in Fig.~\ref{fig xbar1} can only modulate an incoming amplitude in range $[-1, 1]$, the corresponding matrix elements must also satisfy $|a_{ij}|\le1$, which requires normalizing $W$ by $\max_{ij}|W_{ij}|$. The additional multiplier $n$ arises from fan-out and fan-in elements both contributing $\sqrt{n}$. These elements are only shown schematically in Fig.~\ref{fig xbar1}, and can be realized as e.g. a binary tree of 1:1 beam splitters or a sequence of 1:k beam splitters with $k$ ranging from $2$ to $n$, the exact implementation is not important. Since fan-out element must distribute the incoming amplitude $x_i$ equally among $n$ channels, unitarity requires that each channel has amplitude $x_i/\sqrt{n}$. Similarly, a unitary fan-in (see Fig.~\ref{fig fanin}) can not map amplitudes $(x_1,\dots, x_n)$ directly to their sum $\sum_i x_i$, but only to its rescaled version $\sum_i x_i/ c_n$ and $c_n = \sqrt{n}$ is the least possible rescaling factor compatible with unitarity.

The key fact here is that a programming algorithm that embeds $W/\alpha_{\max}(W)$ can not be used to perform the SVT. Indeed,  $\| W \|_{\max}=\max_{ij} |W_{ij}|$ is a type of matrix norm, and it satisfies $\| W \|_{\max} \ge \sigma_{\max}/n$ (see App.~\ref{app norms}), so that for any matrix $W$ it holds
\begin{align}
    \sigma_{\max}(W)\le \alpha_{\max}(W) \ . \label{max inequality}
\end{align}
Therefore, the programming algorithm for the coherent crossbar architecture avoids the SVT argument because it can only embed rescalings $A\sim W$ guaranteed to satisfy $\sigma_{\max}(A)\le 1$.

The trade-off here is that $\alpha_{\max}$ can be significantly larger than $\sigma_{\max}$ for some matrices, in fact (see App.~\ref{app norms})
\begin{align}
1\le \frac{\alpha_{\max}(W)}{\sigma_{\max}(W)}\le n \ .
\end{align}
The higher this ratio, the less energy arrives at the output channels compared to the optimal unitary embedding. For crossbar in Fig.~\ref{fig xbar1} there can be up to $n^2$ less energy compared e.g. to the Clements-based interferometer, which ultimately limits the signal-to-noise ratio of the matrix multiplication.

Inspecting Fig.~\ref{fig xbar1} we see that part of the energy is lost at each MZI block (every time when the multiplication by $|a_{ij}|<1$ is performed). This can actually be fixed in a modified crossbar scheme shown in Fig.~\ref{fig xbar2}. Here, energy from the second output port of each MZI is redirected to the next MZI block rather than being lost. Though this requires to adjust the parameters in each row based on the parameters of the previous row ($a_{ki}$ depend on $a_{(k-1)i}$), this is a small overhead and the overall programming cost is still quadratic. Now the programming is possible as long as $\sum_j a_{ij}^2\le 1$ for each $i$, which implies that the total energy in each row does not exceed the input energy in the corresponding channel. Therefore, interferometer in Fig.~\ref{fig xbar2} can embed the following rescaling of an arbitrary matrix $W$
\begin{align}
    A = \frac{W}{\alpha_{1, 2}(W)}, \quad \alpha_{1, 2}(W)=\sqrt{n}\| W\|_{1, 2} \ .
\end{align}
Here $\| W\|_{1, 2} = \sqrt{\max_i \left(\sum_j W_{ij}^2\right)}$ is another common matrix norm, and it satisfies $\| W\|_{1, 2}\ge \sigma_{\max}(A)/\sqrt{n}$ (see App.~\ref{app norms}). Therefore, $\alpha_{1,2}(W) \ge \sigma_{\max}(W)$ and $A$ is guaranteed to have $\sigma_{\max}(A)\le 1$. Now, however, there is a tighter relation to the optimal rescaling
\begin{align}
1\le \frac{\alpha_{1, 2}(W)}{\sigma_{\max}(W)}\le \sqrt{n} \ ,
\end{align}
meaning that the useful output energy can be at most $n$ times lower compared to the optimal embedding.

We thus find an inevitable trade-off here. Either, a programming algorithm can implement an arbitrary matrix $A=W/\sigma_{\max}(W)$ yielding optimal output energy and has high numerical complexity. Or it can only embed a rescaling $A=W/\alpha(W)$ for which the property $\sigma_{\max}(A)\le 1$ can be checked in $O(n^2)$. However, such rescalings $\alpha(W)$ are in general suboptimal. To the best of our knowledge, all upper bounds $\alpha(W)$ on the largest singular value that can be computed in $O(n^2)$ satisfy
\begin{align}
    \max_W \frac{\alpha(W)}{\sigma_{\max}(W)}=\sqrt{n} \ .
\end{align}
In this sense, it is not possible to asymptotically improve energy efficiency of the scheme in Fig.~\ref{fig xbar2}.

\section{Discussion}
The main observation of this paper is that many optical matrix-vector multiplier architectures have high intrinsic complexity of programming, while those that do not are generally less energy efficient, yielding up to $n$ times less useful output energy.

First, let us discuss possible generalizations of these results. While our exposition and examples focused on the integrated photonic interferometers, the main guiding principle was unitarity, which is applicable much more broadly. For instance, free-space optical multiplication schemes \cite{Spall2020} should share the same trade-offs. 

Next, we anticipate that an even stronger version of our observation might hold. We argued that modifying the matrix to be embedded can make the programming simpler, at the cost of energy inefficiency. However, we only considered rescaling by a scalar value $A=W/\alpha(W)$. What if one allows more general transformations $A=A(W)$ that (i) have quadratic complexity and (ii) ensure that $\sigma_{\max}(A)\le1$? One example is to multiply $W$ by a diagonal matrix $A=W\Lambda^{-1}$. Note that this transformation is easy to compensate by rescaling the inputs $x\to \Lambda x$. For instance, for interferometer in Fig.~\ref{fig xbar2} we can choose $\Lambda_i = \sum_{j} w_{ij}^2$ instead of $\alpha = \max_i\Lambda_i$. This will somewhat improve energy efficiency, because all energy entering each row will now be fully distributed to summators (no energy will reach the red ports in Fig.~\ref{fig xbar2}). The summators themselves, however, will still lead to energy damping up to a factor of $n$. It seems plausible that our argument can be generalized to a broader class of encoding strategies.

We should also mention a related but different task of programming unitary interferometers to implement \textit{unitary} matrices. For instance, the Clements or Reck mesh can be programmed to implement any unitary. Note that this programming has non-trivial complexity as well, e.g. for the Clements mesh it is equivalent to the QR-decomposition \cite{Cilluffo2024}, which has superquadratic complexity \cite{Knight1995}. This complexity is intrinsic for the simple reason that any such programming algorithm indirectly tests whether the input matrix is actually unitary, and apparently there is no deterministic test for unitarity with quadratic complexity.

Now we list a number of limitations and implicit assumptions of our work. First, the core argument relied on the complexity of certain linear algebraic primitives, e.g. the singular value test. While we believe it is implausible that any of these can take significantly less than $O(n^3)$ in practice, our knowledge of the relevant mathematical results may be incomplete.

We also assumed that these primitives must be exact, deterministic, and applicable to arbitrary input matrices. Relaxing any of these assumptions may allow for algorithms with quadratic complexity. For instance, checking if a matrix is unitary can be performed in $O(n^2)$ with probability exponentially close to one (it suffices to check whether the norm of a random vector is conserved). Hence, a programming algorithm that  is either approximate, randomized, or valid for a subclass of matrices only, can in principle have quadratic complexity. 

Finally, we assumed a very specific way of encoding the target linear transformation $W$ into the unitary matrix, i.e. that input $x$ and output $y$ vectors are channel amplitudes. Our argument does not directly apply to more general encodings using e.g. intensity or phase information. 

\appendix   
\section{Unitary embedding} \label{app unitarity}

A matrix $A$ can be embedded in a unitary $U$ as in \eqref{embedding}, if the unitarity equations $U^\dagger U=\mathbb{I}$ can be solved for $B, C, D$. Explicitly,
\begin{align}
&A^\dagger A + C^\dagger C = \mathbb{I}_{n\times n}\\
&A^\dagger B+C^\dagger D=0_{n\times m}\\
&B^\dagger B+D^\dagger D = \mathbb{I}_{m\times m }
\end{align}

Consider the SVD of matrix $A=V \Sigma W$. The first equation can be rewritten as
\begin{align}
C^\dagger C=W(\mathbb{I}-\Sigma^2)W^\dagger   \label{eq C}\ .
\end{align}
Because $C^\dagger C$ is positive-semidefinite, this equation can only be solved if all singular values of $A$ are less or equal to one $\sigma_i(A)\le 1$. If this condition holds, \eqref{eq C} can be solved as long as $C$ has sufficient rank. Generically, the rank of $\mathbb{I}_{n\times n}-\Sigma^2$ is $n$ and the rank of $C^\dagger C$ is $\min(n, m)$, so as long as $m\ge n$ \eqref{eq C} can be solved, e.g. one can choose $C=V\sqrt{\mathbb{I}-\Sigma^2} W$. Blocks $B$ and $D$ can be found similarly. A possible embedding of $A$ reads (note that the solution is not unique)
\begin{align}
    U = \begin{pmatrix} V \Sigma W & V\sqrt{1-\Sigma^2}W \\ V \sqrt{1-\Sigma^2}W & -V\Sigma W\end{pmatrix} \ .
\end{align}

\section{Matrix norms} \label{app norms}
The Frobenius norm can be expressed as the sum of the singular values squared
\begin{align}
    \|W\|_F^2=\operatorname{Tr}W^2=\sum_{ij}w_{ij}^2 = \sum_i \sigma_i^2 \ .
\end{align}
Now, to show that $\|W\|_{\max}=\max_{ij} |w_{ij}|$ satisfies $\|W\|_{\max}\ge \sigma_{\max}/n$ we construct a simple series of inequalities
\begin{align}
    n^2\|W\|^2_{\max} \ge \sum_{ij} w_{ij}^2 = \sum_i \sigma_i^2\ge \sigma_{\max}^2 \ .
\end{align}
Similarly, we can write
\begin{align}
    n\max_i \sum_j w_{ij}^2\ge \sum_{ij} w_{ij}^2=\sum_i \sigma_i^2\ge \sigma_{\max}^2 \ .
\end{align}
Hence, $\|W\|_{1,2}=\sqrt{\max_i \sum_j w_{ij}^2}$ satisfies $\|W\|_{1,2}\ge \sigma_{\max}/\sqrt{n}$.
\bibliographystyle{quantum}
\bibliography{library.bib}

\end{document}